# Effect of Magnetic-Field-Induced Restructuring on the Elastic Properties of Magnetoactive Elastomers

Andrei A. Snarskii[1,2,*], Mikhail Shamonin[3,*], Pavel Yuskevich[1],

[1] *National Technical University of Ukraine "Igor Sikorsky Kyiv Polytechnic Institute", Prospekt Peremohy 37, 03056 Kiev, Ukraine*

[2] *Institute for Information Recording, NAS of Ukraine, Mykoly Shpaka Street 2, 03113 Kiev, Ukraine*

[3] *East Bavarian Centre for Intelligent Materials (EBACIM), Ostbayerische Technische Hochschule (OTH) Regensburg, Seybothstr. 2, 93053 Regensburg, Germany*

**Abstract**

Composite materials where magnetic micrometer-sized particles are embedded into a compliant polymer matrix are known as magnetorheological or magnetoactive elastomers (MAEs). They are distinguished by huge variations of their physical properties in a magnetic field, which is commonly attributed to the restructuring of the filler. The process of the magnetic-field-induced restructuring in a magnetorheological elastomer is interpreted as progression towards percolation. Such a physical model was previously used to explain the dependence of the magnetic permeability and dielectric permittivity of MAEs on the magnetic field strength. Based on this hypothesis, the magnetorheological effect in MAEs is considered theoretically. The theoretical approach is built upon a self-consistent effective-medium theory for the elastic properties, extended to the variable (field dependent) percolation threshold. The proposed model allows one to describe the large variations (over several orders of magnitude) of the effective elastic moduli of these composite materials, known as the giant magnetorheological (MR) and field-stiffening effects. An existence of a giant magnetic Poisson effect is predicted. The relation of the proposed model to the existing theories of the MR effect in MAEs is discussed. The results can be useful for applications of MAEs in magnetic-field controlled vibration dampers and isolators.

[*] Correspondence: asnarskii@gmail.com (A.S.); mikhail.chamonine@oth-regensburg.de (M.S.)

## 1. Introduction

Magnetoactive elastomers (MAEs) are a class of composite materials where significant changes of physical properties are observed in moderate (several hundred mT) dc magnetic fields [1-9]. The most prominent phenomenon is the magnetorheological (MR) or field-stiffening effect, when the elastic moduli of these materials grow over several (up to four) orders of magnitude, respectively. In this context, these materials are often referred to as magnetorheological elastomers. MAEs consist of rigid, ferromagnetic magnetic particles (usually of a spherical shape) dispersed in a compliant (soft) polymer matrix. It is essential that the elastic moduli of the inclusions (e.g. iron, shear modulus $G_{Fe} \sim 10^{11}$ Pa) are many orders of magnitude larger than those of the matrix (e.g. polydimethylsiloxane (PDMS) $G_{PDMS} \sim 10^3 - 10^5$ Pa).

Large theoretical efforts have been made for explaining the giant [10] (or even colossal [11]) MR effect in MAEs [12,13]. The general approaches can be roughly divided into two different groups: i) the cause is in the interaction between induced magnetic moments between the particles [14] (here the simplest model is a system of point dipoles connected by elastic springs [15]); ii) the cause is the rotation of individual particles [16-19]. Although these mechanisms are undoubtedly contribute to the MR effect in MAEs, the existing approaches are not satisfactory when it comes to the description of the MR effect over several (say, three) orders of magnitude. As an example, a 20-fold field-induced increase of the moduli has been calculated in [20]. Hitherto, the existing theoretical approaches concentrate on the magnetic interactions between the inclusions and the self-interaction of a magnetized particle with a magnetic field and largely disregard the effect of the change in the composite's microstructure (i.e. mutual arrangement of filler particles) on its elastic properties. However, it has to be expected that the change in the microstructure should alter the effective elastic properties of a composite material. For example, in [21], a one-dimensional model for ferrogels was considered, where "hardened" states with touching particles and therefore diverging compressive elastic modulus were found.

The purpose of this paper is to propose a complementary physical model based on the idea that the restructuring (RS) of the filler [22-24] means the progression towards the percolation structure. Figure 1 illustrates the proposed physical picture [25].

When a magnetic field (designated by an arrow) is applied to the composite, a significant part of the magnetic flux passes through the pre-cluster (shaded region). In Figure 1, only a small part of the structure of the cluster is drawn, which in fact does not represent a column, but a complex winding line (cf. Figure 5.5 on page 97 of [26]).

The ferromagnetic particles in a vicinity of the pre-cluster are "sucked" towards pre-cluster by magnetic forces. A visual representation of such an effect, obtained by means of X-ray tomography, was presented in [27]. In [28], a virtual touching and detachment of rigid inclusions in a soft elastic matrix by employment and elimination of a magnetic field was observed. This can be considered as an elementary step towards the build-up (or the destruction) of an infinite cluster in magnetic field (or its removal).

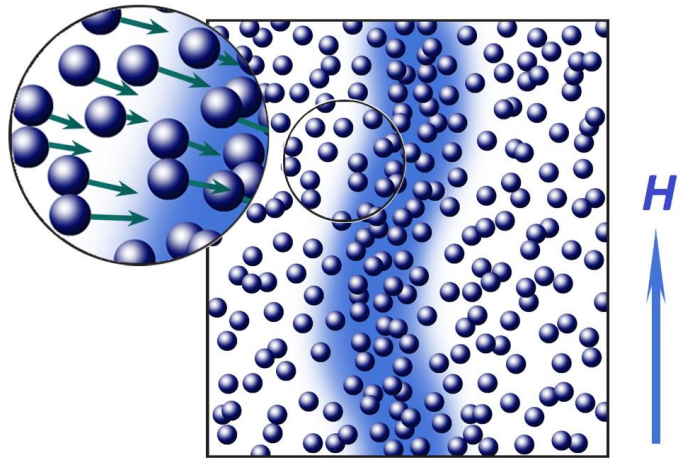

**Figure 1**. Artist's impression of the restructuring in an applied magnetic field.

Such a reconfiguration of the microstructure is quantitatively characterized by the difference between the total particle concentration $p$ and the field-dependent percolation threshold (PT) $p_c$. Although we do not use the methods of percolation theory [29,30], the modified effective medium theory (EMT [31]) comprehends the existence of the PT (i.e. a particular concentration where there is a steep rise of elastic moduli) and its dependence of an external magnetic field. Given that the concept of progression towards percolation [25] successfully described the magnetodielectric effect and the non-monotonic dependence of the effective magnetic permeability on an external magnetic field in MAEs, it can be expected that the same mechanism is significant for the elastic and other magnetic-field-dependent physical properties of MAEs. It will be shown below that our model is capable of predicting the correct order of magnitude of the MR effect, while keeping the qualitative behavior (increase of elastic moduli with increasing magnetic field) unaltered. As a mathematical instrument, we will employ the self-consistent EMT for the elastic properties, modified in such a way that a steep change in an elastic modulus occurs at a given concentration of rigid inclusions. We utilize the previously introduced concept of the moveable (field dependent) PT [25] and use the empirical dependence of the PT on a magnetic field found in [32].

Recall that in analogy to MR fluids, MR effect in MAEs is often ascribed to the formation of chainlike aggregates along the magnetic field lines. This simplified physical picture for high concentrations of magnetic particles was recently questioned by Romeis *et al.* [33], whose numerical simulations showed that development of elongated structures becomes impossible due to purely geometrical constraints. The proposed physical mechanism of the MR effect in highly filled MAEs would be indirectly supported by these calculations: small movements of magnetized particles within a MAE specimen may result in drastic changes of its magnetorheological properties.

The paper is organized as follows: In the next Section the two-phase composite material of interest is described and a modified self-consistent EMT for the elastic properties of two-phase composite materials, with the included concept of a moveable percolation threshold [34], is presented. The results of calculations are described in Section 3 and discussed in Section 4. Conclusions are drawn in the final section.

2. **Materials and Methods.**

In the following, we consider random heterogeneous two-phase composites. The self-consistent EMT approximation considers inclusions of the spherical shape embedded into a fictitious homogeneous medium with the effective elastic properties searched. The concentration of the first phase is $p$, the concentration of the second phase is ($1 - p$). The mechanical properties of both phases and the effective medium are isotropic.

$G_e, \nu_e$ denote the effective shear modulus and the Poisson's ratio, respectively, while $G_1, G_2, \nu_1, \nu_2$ are the values of these moduli in the first and second phases.

To be specific, we assume iron ($G_1 = 80.7\,\text{GPa}$, $\nu_1 = 0.3$) as a ferromagnetic filler (first phase) embedded in a soft polydimethylsiloxane matrix (second phase; $G_2 = 7.43\,\text{kPa}$, $\nu_2 = 0.49$). With these materials, the volume concentration $p = 0.23$ corresponds to the typical concentration of iron particles of about 70 mass%. The calculated effective shear modulus at $p \approx 0.23$ in the absence of a magnetic field is $G_e(0) \approx 40\,\text{kPa}$ [35]. The material is isotropic (unstructured) in the absence of a magnetic field.

The system of equations of the classical self-consistent EMT [36,37] can be written in the following form:

$$\left. \begin{array}{l} \Omega_1 p + \Omega_2 (1-p) = 0 \\ \Theta_1 p + \Theta_2 (1-p) = 0 \end{array} \right\}, \qquad (1)$$

where

$$\Omega_i = \frac{\dfrac{G_i}{G_e} \cdot \dfrac{1+\nu_i}{1+\nu_e} \cdot \dfrac{1-2\nu_e}{1-2\nu_i} - 1}{1 + \alpha_e \left( \dfrac{G_i}{G_e} \cdot \dfrac{1+\nu_i}{1+\nu_e} \cdot \dfrac{1-2\nu_e}{1-2\nu_i} - 1 \right)}, \; \Theta_i = \frac{\dfrac{G_i}{G_e} - 1}{1 + \beta_e \left( \dfrac{G_i}{G_e} - 1 \right)}, \; \alpha_e = \frac{1}{3} \cdot \frac{1+\nu_e}{1-\nu_e}, \; \beta_e = \frac{2}{15} \cdot \frac{4-5\nu_e}{1-\nu_e}. \qquad (2)$$

Very recently, it was proposed to modify the equations of EMT for the elasticity problem in the following way [34]:

$$\left. \begin{array}{l} \dfrac{\Omega_1}{1+s(p,\tilde{p}_c)\Omega_1} p + \dfrac{\Omega_2}{1+s(p,\tilde{p}_c)\Omega_2}(1-p) = 0 \\ \dfrac{\Theta_1}{1+s(p,\tilde{p}_c)\Theta_1} p + \dfrac{\Theta_2}{1+s(p,\tilde{p}_c)\Theta_2}(1-p) = 0 \end{array} \right\}. \qquad (3)$$

Similar modification was established for galvanomagnetic phenomena in [38].

The term $s(p, \tilde{p}_c)$ is

$$s(p, \tilde{p}_c) = (1 - 2\tilde{p}_c) \left(\frac{p}{\tilde{p}_c}\right)^{\tilde{p}_c} \left(\frac{1-p}{1-\tilde{p}_c}\right)^{1-\tilde{p}_c}. \qquad (4)$$

It allows one to set the PT of the first phase $p_c$ to $\tilde{p}_c$.

In [30], the following empirical relationship was proposed:

$$\tilde{p}_c\left(|\langle \mathbf{H} \rangle|\right) = \tilde{p}_c(0) e^{-\frac{|\langle \mathbf{H} \rangle|}{H_c}}, \qquad (5)$$

where **H** is the magnetic field inside the composite material, $H_c$ is the characteristic magnetic field strength, $\langle ... \rangle = 1/V \int ... dV$, $V$ is the averaging volume, wherein the characteristic dimensions of the averaging region should be much larger than the correlation length. In the following, we denote $|\langle \mathbf{H} \rangle|$ as $H$. The order of magnitude of $H_c$ was found to be $10^5 - 10^6$ A/m [25,32]. To the best of our knowledge, the hypothesis of the field dependence of the percolation threshold, for the case of magnetorheological fluids, was introduced in [39]. The physical meaning of the critical magnetic field $H_c$ is being a characteristic magnetic field at which the significant RS of the filler takes place. If the polymer matrix is so stiff, that the RS may not occur, the critical field should theoretically tend to infinity $H_c \to \infty$. It has to be expected that at the same filler concentration the critical field is smaller for a softer elastomer matrix.

In [32], equation (5) was used in the percolation formula $G_e \sim (p - \tilde{p}_c)^\gamma$, $p > \tilde{p}_c$. However, percolation formulas work only in the very narrow region (critical region) in a vicinity of the PT. In [25], a modified EMT allowed one to consider the entire concentration range up to the PT and provided a good agreement with measurements of dielectric and magnetic properties of MAEs. The concept of a moveable percolation threshold has been extended to the elastic properties of composite materials in [34], where it was shown that this approach is capable of explaining several known experimental results.

The system of transcendental equations (1) is solved numerically, for given values of parameters $p$, $H$, $\tilde{p}_c(0)$, $G_1, G_2, \nu_1, \nu_2$ and $H_c$.

## 3. Results

Figure 2 shows the dependence of the field-dependent PT $\tilde{p}_c(H)$ on the magnetic-field strength $H$ for a number of various values of the critical field $H_c$. The dashed horizontal line designates the value $p = 0.23$ while $\tilde{p}_c(0) = 1/3$. A particular value of the critical magnetic field, $H_c = 620\,\text{kA/m}$, corresponds to the value obtained in [25], where it has been used to describe the magnetic properties of an MAE sample with $p \approx 0.23$ and the shear modulus in the absence of a field $G_e(0) \approx 40\,\text{kPa}$ (reported for this material in [35]) and provided good agreement between theory and experiment. The $\tilde{p}_c(0)$ value of 1/3 is also selected according to the comparison between theoretical and experimental results in [25].

In the region $p < \tilde{p}_c(H)$, the model (1) – (5) describes a composite material with inclusions of the phase 1 embedded into a matrix (phase 2). Above the PT, the equations describe inclusions of the second phase in the first phase, which does not correspond to the MAE structure, although the calculation results can be qualitatively correct. In the following, the solutions above the PT are shown by dashed lines. Vertical dashed lines in Figure 2 indicate the maximum value of the magnetic field $H_{\max}$, where the solution of (4) correspond to the microstructure of interest. It can be easily obtained that $H < H_{\max} = H_c \ln(\tilde{p}_c(0)/p)$.

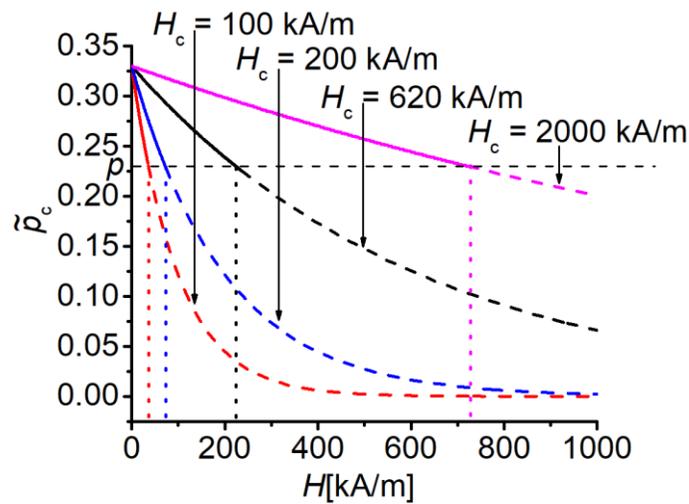

**Figure 2.** Field dependence of the percolation threshold for different critical magnetic fields $H_c$. The horizontal dashed line corresponds to $p = 0.23$.

The impact of the field-dependent behaviour of the PT is presented in Figure 3. At a given filler concentration $p$, the effective shear modulus grows with the increasing magnetic field (Fig. 3 a). Figure 3 b demonstrates that the growth of the effective shear modulus upon nearing towards PT increases with increasing magnetic field. Follow the positions of dots corresponding to the shear modulus at the concentration value $p = 0.23$.

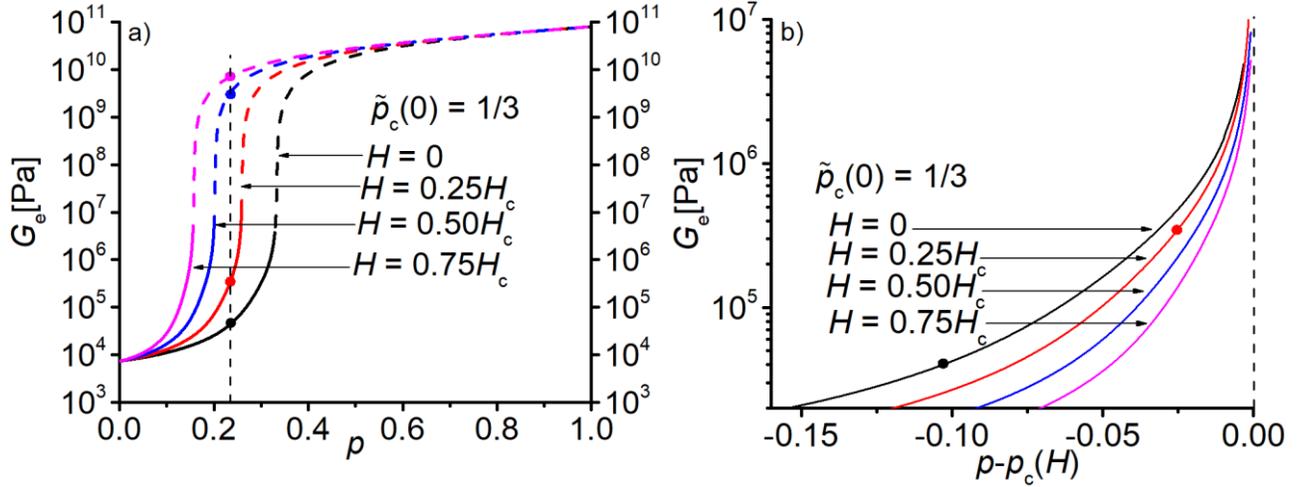

**Figure 3.** (a) The calculated concentration dependences of the effective shear modulus for different applied fields. The vertical dashed line designates $p = 0.23$. The dots denote the value of the shear modulus at $p = 0.23$. (b) The concentration dependences below the field-dependent percolation threshold $p_c(H)$, presented *versus* the "concentration distance" to the percolation threshold, $p - p_c(H)$. The dots denote the value of the shear modulus at $p = 0.23$.

Figure 4 depicts the calculated field dependences of the effective shear modulus and the corresponding MR effect defined as

$$MRE(H) = \frac{G_e(H) - G_e(H=0)}{G_e(H=0)} \ . \tag{6}$$

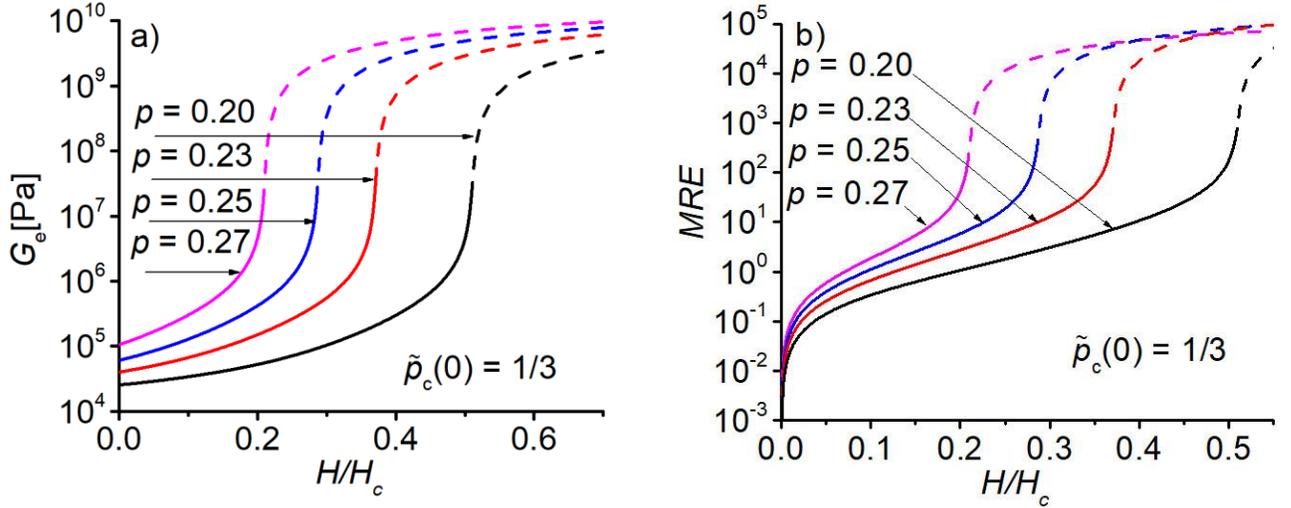

**Figure 4.** Field dependences of the effective shear modulus (a) and the magnetorheological effect (b) for different concentrations of the filler material.

It is observed that both the effective shear modulus and the MR effect grows with an increasing magnetic field and tends to saturate in large magnetic fields. The magnitude of MR effect is bigger for the larger concentration of inclusions. Within the applicability region of the model with respect to the microstructure of the material, the effect reaches the order of magnitude of $10^3$.

### 4. Discussion

The results presented in Figure 4 demonstrate that the MR effect can reach very large values due to changes in the composite's microstructure, caused by magnetic forces.

Moreover, the model predicts a giant variation of the Poisson's ratio with the applied magnetic field (Figure 5). We take courage to name this phenomenon – the magnetic Poisson effect (MPE) and embolden experimentalists to verify our prediction. It is quantified as

$$MPE(H) = \frac{v_e(H) - v_e(H=0)}{v_e(H=0)}. \tag{7}$$

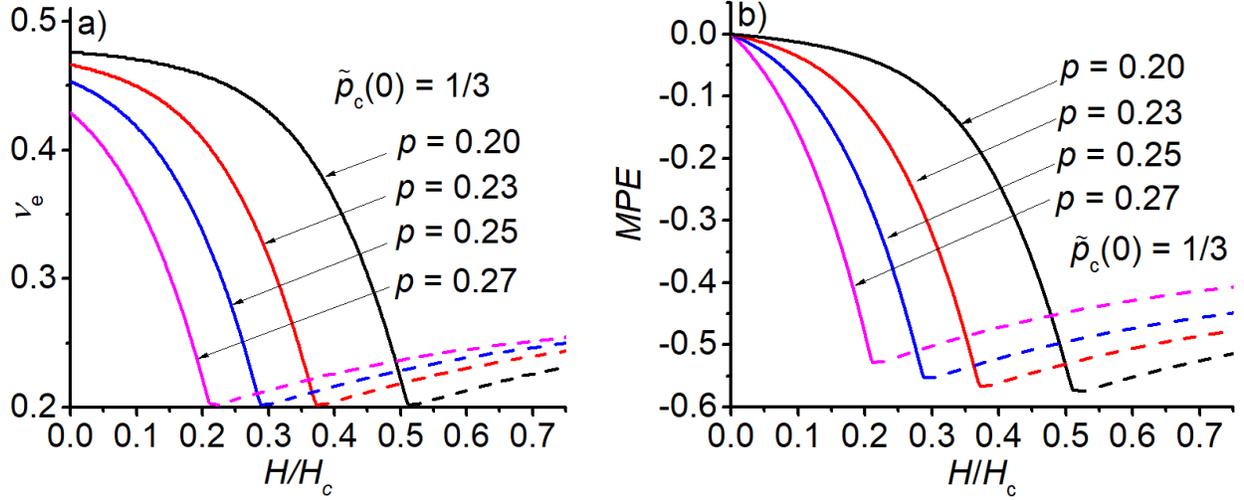

**Figure 5.** Field dependence of the Poisson's ratio (a) and the resulting magnetic Poisson effect (b) for different concentrations of the filler material.

In the framework of the EMT, the effective elastic modulus of a random heterogeneous material can not be smaller than $G_1$ and larger than $G_2$. The effective Poisson's ratio $\nu_e$ behaves in a more peculiar way. In general, it is not limited between $\nu_1$ and $\nu_2$ [40]. For a large inhomogeneity (as in the case of interest), it reaches at the percolation threshold the value close to 1/5, which is smaller than both $\nu_1$ and $\nu_2$ in our case [34]. The particular value of the Poisson's ratio $\nu^* = 1/5$ was discussed in [41]. In this sense, the predicted change in an effective Poisson's ratio can be considered more significant than the progression of $G_e$ towards $G_1$ and, therefore, qualified as a "giant" effect. An experimental search for the change of an effective Poisson's ratio in magnetic fields can be considered as a way to verification of the proposed mechanism of the MRE.

It is worth noting that, for a given critical magnetic field $H_c$ the field dependence of the effective shear modulus can be reduced approximately to a single "master curve". This is demonstrated in Figure 6 a, where the effective shear modulus for four different filler concentrations $p = 0.20, 0.23, 0.25, 0.27$ is presented as a function of the "field distance" to the PT, $H_{max} - H$. These three curves are indistinguishable on the scale of Fig. 6 a. A closer look at the field dependence (Figure 6 b) reveals that some difference is, however, present. At percolation threshold ($H = H_{max}$), the value of $G_e \approx 3.7 \cdot 10^7$ Pa is reached. In zero magnetic field, at a given filler concentration, the composite material has a specific shear modulus, which grows with the increasing concentration $p$. This is the other limiting point in Figure 6, where each curve ends.

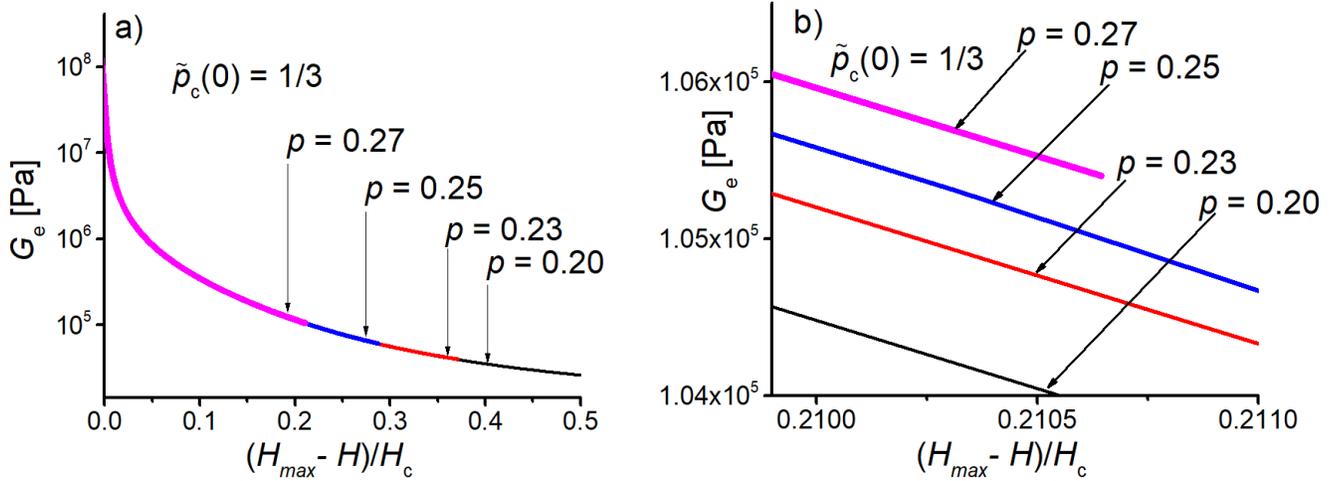

**Figure 6.** Normalized-field dependence of the effective shear modulus on the large scale of field variation (a) and on the small scale of field variation (b).

We excluded other possible mechanisms (magnetic forces between magnetized particles, rotations of particles due to an applied magnetic field) from our consideration. The calculated effect is solely due to RS of the filler. The order of magnitude of the colossal MR effect in ultra-soft MAEs can be explained. In [11], the maximum MR effect was observed for the MAE material with $p \approx 0.26$ (75 mass% of iron) and $G_e(0) \approx 273$ Pa. Such a value of the effective shear modulus $G_e(0)$ is theoretically achieved with $\tilde{p}_c(0) = 1/3$ and $G_2 \approx 25.5$ Pa, which was reported in [11] as the lowest obtained shear modulus of the polymer matrix. Substituting these material parameters in equations (1)-(5), we obtain for the parameter MRE close to the percolation threshold $H \to H_{max}$ the following value: MRE $\approx 0.8 \times 10^4$, which describes the order of magnitude (~ $10^4$) of the measured MRE well.

Figure 4 predicts that the maximum achievable value of the shear modulus (where the composite morphology correspond to that of an original material – inclusions of the first phase embedded into the matrix of the second phase) should be of the order of magnitude of several tens of MPa. Conventional models based on the concept of magnetic interactions between particles in chain-like structures estimate the maximum value of the magnetic-field induced shear modulus of about 0.5 MPa [23,42]. In experiments (see e.g. [11]), the maximum shear modulus of several MPa can be observed and the value of about ten MPa is predicted at magnetic saturation by extrapolation of experimental results. Our theory seems to be closer to the real maximum values than previously proposed approaches. In the framework of the EMT theory, it should be possible to reach the magnitude of the magnetic-field-induced shear modulus comparable with the elastic modulus of iron, which is significantly larger (~ 100 GPa). Why is such a magnitude of the shear modulus not observable in experiment? This is explained by the fact that above the percolation threshold, the self-consistent EMT cannot describe inclusions of the first phase embedded into the second phase. In EMT, the self-consistency of fields is considered: The mechanical stress in the isolated inclusion of the first phase in the effective medium and stress in the isolated inclusion of the second phase in the effective medium compensate each other and result in a stress of the effective medium. This means that at small concentrations, such an approximation gives a description of the inclusions of the first phase in the matrix of the second phase. At large concentrations, it describes the effective properties of the inclusions of the second phase in the matrix of the first phase. At the same time, owing to fabrication technology, our composite material at any concentration represents spherical inclusions of the first phase (iron) in the continuous second phase (polymer matrix) and never can be described as discrete inclusions of the second phase in the continuous first phase, i.e. spherical polymer particles in the iron matrix. In general, in composites fabricated using other technologies and compositions, the morphology could be different.

Why is the observed maximal value of the magnetic-field induced shear modulus typically few times less than the value predicted by the EMT just below the percolation threshold? We see one reason in the inaccuracy of the EMT approximation. In the EMT or the percolation theory for electro-conductive (or dielectric) properties, the touching of well conductive particles in a badly conductive means formation of the well conductive path. In real situation, when measuring the shear modulus of a composite material, the particles do not form a rigid mechanical bond to each other (they can slip relative to each other). This would result into a reduction of an effective shear modulus.

Figure 4 also shows that large changes in the effective shear modulus due to RS of the filler occur only in the vicinity of the percolation threshold $\tilde{p}_c(H)$, i.e. when $H \sim H_{max}$. If $H \ll H_{max}$, these changes are moderate (MRE $\leq 10^2$). In the proposed EMT, the effective shear modulus $G_e$ is a concave function of the magnetic field $H$ (typical sigmoid curves are observed on the logarithmic scale in Figure 4), where a steep rise of $G_e$ occurs when $H \to H_{max}$. In experiments, the situation is somewhat different. Usually, the field dependence of $G_e$ is concave only for small magnetic fields $H \approx 0$, then, after an inflection point, the function $G_e(H)$ becomes convex, showing the tendency to saturation $\left(\lim_{H \to \infty} G_e(H) = G_e^{sat}\right)$. The complete saturation of $G_e(H)$ is usually not observable in experiments due to the large demagnetizing factor of thin samples. In terms of the external magnetic flux density, such a dependence of the change in the shear storage modulus has been empirically expressed by Zrínyi et al. [43] as

$$\Delta G_e = G_e^{sat} \times \frac{B^2}{a_B^2 + B^2}. \tag{8}$$

Typically $a_B \sim 0.3$ T [11], which corresponds to the magnetic field strength of approximately $2.4 \times 10^5$ A/m in the air. It can be seen that, by the order of magnitude, this value agrees well with the values of the critical magnetic field $H_c$ obtained in [25,32]. Recall that conventional theories of MRE underestimate the value of $G_e^{sat}$ in ultra-soft ($G_e$ in the absence of a magnetic field is roughly below 10 kPa) MAEs. This can be rectified by the presented theory. The saturation behavior of the magnetic-field-induced shear modulus occurs naturally in theoretical models considering interactions between magnetized inclusions (usually simplified as point dipoles, which is definitely not true if spherical particles are close to each other) [44] or where the field-induced uniaxial magnetic anisotropy counteracts the shear deformation [45]. The reason is the saturation of particle magnetization with the increasing magnetic field [46]. The presented theoretical model does not exclude other possible mechanisms, which can take place simultaneously but become more or less significant at different levels of magnetization.

Previously, it has been shown that the EMT with moveable percolation threshold is capable of describing the magnetic and magnetodielectric effects in MAEs qualitatively and quantitatively [25]. As far as the magnetic or magnetodielectric properties are concerned, they are determined by the mutual arrangement of inclusions in an MAE. It is irrelevant if there are magnetic forces or moments acting on the particles, although, of course, these forces and moments led to the variation of the microstructure. Moreover, during the magnetic or magnetodielectric measurements the shape of an MAE sample is fixed, the internal microstructure varies only due to a magnetic-field.

In elastic measurements, an additional effect on the microstructure occurs because of external mechanical forces. During magnetorheological measurements, the specimen is mechanically deformed in order to probe the material rigidity. Obviously, such a mechanical loading tries to "tear off" the particle network and the magnetic interactions become important. A prominent example is the magnetic-field-enhanced Payne effect, which can be observed in MAEs [47,48].

## 5. Conclusion and Outlook

Hitherto, the concept of the field-dependent percolation threshold allowed one to describe the magnetodielectric effect and the non-monotonous field dependence of the magnetic permeability in MAEs [25]. The calculations in the present paper show that the same physical mechanism gives the previously unexplained order of magnitude for the giant or colossal MR effect. Therefore, it has to be taken into account when considering the MR effect in MAEs. This model predicts a significant change in a Poisson's ratio of compliant MAEs in external magnetic fields. It is proposed to use the measurement of a Poisson's ratio as a verification test for this theoretical model.

Our model does not exclude alternative mechanisms, which may be present simultaneously and should also contribute to the field-stiffening or magnetorheological effects (by further enhancement).

The application of an external constant magnetic field should induce anisotropy, which has to be taken into account in the calculation of effective elastic properties. For magnetic properties, the presence of the uniaxial anisotropy was proven experimentally in [35]. In the present model, it is assumed that the anisotropy is small. We hypothesize that the anisotropy can be introduced into the effective medium approximation by assuming two different percolation thresholds in two directions, parallel and perpendicular to an applied magnetic field.

The advantage of the proposed model is that it allows one to describe in a unified manner magnetic properties, magnetodielectric effect [25] and, henceforth, to explain the mechanism of the giant increase in elastic properties of MAEs. All physical properties should originate from one and the same arrangement of inclusions in a composite material. If a percolating structure comes into play in a composite material, its existence must be seen in several physical properties (cross-property relations). An important next step should be theoretical explanation of the empirical rule (5) proposed in [32]. In particular, the relation of the critical magnetic field $H_c$ to the physical properties of a composite material and its constitutive components has to be established.

## References

[1] G. Filipcsei, I. Csetneki, A. Szilágyi and M. Zrínyi, "Magnetic field-responsive smart polymer composites," in *Oligomers-Polymer Composites-Molecular Imprinting*, Berlin, Heidelberg, Springer, 2007, pp. 137-189.

[2] Y. Li, J. Li, W. Li and H. Du, "A state-of-the-art review on magnetorheological elastomer devices," *Smart. Mater. Struct.,* vol. 23, no. 12, 123001, 2014.

[3] S. Odenbach, "Microstructure and rheology of magnetic hybrid materials," *Arch. Appl. Mech.,* vol. 86, no. 1-2, pp. 269-279, 2016.

[4] A. M. Menzel, "Tuned, driven, and active soft matter," *Phys. Rep.,* vol. 554, pp. 1-45, 2015.

[5] Ubaidillah, J. Sutrisno, A. Purwanto and S. A. Mazlan, "Recent progress on magnetorheological solids: materials, fabrication, testing, and applications," *Adv. Eng. Mater.,* vol. 17, no. 5, pp. 563-597, 2015.

[6] M. T. Lopez-Lopez, J. D. Durán, L. Y. Iskakova and A. Y. Zubarev, "Mechanics of magnetopolymer composites: a review.," *J. Nanofluids,* vol. 5, no. 4, pp. 479-495, 2016.

[7] M. A. Cantera, M. Behrooz, R. F. Gibson and F. Gordaninejad, "Modeling of magneto-mechanical response of magnetorheological elastomers (MRE) and MRE-based systems: a

review," *Smart Mater. Struct.,* vol. 26, no. 2, 023001, 2017.

[8] R. Weeber, M. Hermes, A. M. Schmidt and C. Holm, "Polymer architecture of magnetic gels: A review," *J. Phys.: Cond. Matt.,* vol. 30, no. 6, 063002, 2018.

[9] M. Shamonin and E. Y. Kramarenko, "Highly Responsive Magnetoactive Elastomers," in *Novel Magnetic Nanostructures. Unique Properties and Applications*, Amsterdam, Elsevier, 2018, pp. 221-245.

[10] A. V. Chertovich, G. V. Stepanov, E. Y. Kramarenko and A. R. Khokhlov, "New composite elastomers with giant magnetic response," *Macromol. Mater. Eng.,* vol. 295, no. 4, pp. 336-341, 2010.

[11] A. Stoll, M. Mayer, G. J. Monkman and M. Shamonin, "Evaluation of highly compliant magneto-active elastomers with colossal magnetorheological response," *J. Appl. Polym. Science,* vol. 131, no. 2, 39793, 2014.

[12] P. Cremer, *Mesoscale Modeling of Magnetic Elastomers and Gels-Theory and Simulations,* Doctoral dissertation, Universitäts-und Landesbibliothek der Heinrich-Heine-Universität Düsseldorf, 2017.

[13] Y. Han, W. Hong and L. E. Faidley, "Field-stiffening effect of magneto-rheological elastomers," *Int. J. Solids Struct.,* vol. 50, no. 14-15, pp. 2281-2288, 2013.

[14] D. Romeis, P. Metsch, M. Kästner and M. Saphiannikova, "Theoretical models for magneto-sensitive elastomers: a comparison between continuum and dipole approaches," *Phys. Rev. E,* vol. 95, no. 4, 042501, 2017.

[15] P. A. Sánchez, O. V. Stolbov, S. S. Kantorovich and Y. L. Raikher, "Modeling the magnetostriction effect in elastomers with magnetically soft and hard particles," *Soft Matter,* vol. 15, no. 36, pp. 7145-7158, 2019.

[16] V. M. Kalita, A. A. Snarskii, M. Shamonin and D. Zorinets, "Effect of single-particle magnetostriction on the shear modulus of compliant magnetoactive elastomers," *Phys. Rev. E,* vol. 95, no. 3, 032503, 2017.

[17] A. A. Snarskii, V. M. Kalita and M. Shamonin, " Renormalization of the critical exponent for the shear modulus of magnetoactive elastomers," *Scientific Reports,* vol. 8, no. 1, pp. 1-8, 2018.

[18] M. V. Vaganov, D. Y. Borin, S. Odenbach and Y. L. Raikher, "Modeling the magnetomechanical behavior of a multigrain magnetic particle in an elastic environment," *Soft Matter,* vol. 15, no. 24, pp. 4947-4960, 2019.

[19] V. M. Kalita, Y. I. Dzhezherya and G. G. Levchenko, "The loss of mechanical stability and the critical magnetization of a ferromagnetic particle in an elastomer," *Soft Matter,* vol. 15,

no. 29, pp. 5987-5994, 2019.

[20] D. Ivaneyko, V. Toshchevikov and M. Saphiannikova, "Dynamic-mechanical behaviour of anisotropic magneto-sensitive elastomers," *Polymer,* vol. 147, pp. 95-107, 2018.

[21] M. A. Annunziata, A. M. Menzel and H. Löwen, "Hardening transition in a one-dimensional model for ferrogels," *J. Chem. Phys. ,* vol. 138, no. 20, 204906, 2013.

[22] N. Felici, J. N. Foulc and P. Atten, "A conduction model of electrorheological effect. Electrorheological Fluids," in *Electrorheological Fluids: Mechanisms, Properties, Technology and Applications*, R. Tao and D. Roy, Eds., Singapore, World Scientific, 1994, pp. 139-152.

[23] M. R. Jolly, J. D. Carlson and B. C. Munoz, "A model of the behaviour of magnetorheological materials," *Smart Mater. Struct.,* vol. 5, no. 5, p. 607, 1996.

[24] E. Y. Kramarenko, A. V. Chertovich, G. V. Stepanov, A. S. Semisalova, L. A. Makarova, N. S. Perov and A. R. Khokhlov, "Magnetic and viscoelastic response of elastomers with hard magnetic filler," *Smart. Mater. Struct.,* vol. 24, no. 3, 035502, 2015.

[25] A. A. Snarskii, D. Zorinets, M. Shamonin and V. V. Kalita, "Theoretical method for calculation of effective properties of composite materials with reconfigurable microstructure: electric and magnetic phenomena," *Phys. A: Stat. Mech. Appl.,* vol. 535, 122467, 2019.

[26] B. I. Shklovskii and A. L. Efros, *Electronic properties of doped semiconductors*, Berlin: Springer Science & Business Media, 2013.

[27] T. Gundermann, P. Cremer, H. Löwen, A. M. Menzel and S. Odenbach, " Statistical analysis of magnetically soft particles in magnetorheological elastomers," *Smart Mater. Struct.,* vol. 26, no. 4, 045012, 2017.

[28] M. Puljiz, S. Huang, K. A. Kalina, J. Nowak, S. Odenbach, M. Kästner, G. Auernhammer and A. M. Menzel, "Reversible magnetomechanical collapse: virtual touching and detachment of rigid inclusions in a soft elastic matrix," *Soft Matter,* vol. 14, no. 33, pp. 6809-6821, 2018.

[29] D. Stauffer and A. Aharony, *Introduction to percolation theory*, London, UK: Taylor & Francis, 2018.

[30] A. Snarskii, I. V. Bezsudnov, V. A. Sevryukov, A. Mozovskiy and J. Malinsky, *Transport Processes in Macroscopically Disordered Media. From Mean Field Theory to Percolation*, New York: Springer Verlag, 2016.

[31] T. C. Choy, *Effective medium theory: principles and applications*. Oxford, UK: Oxford

University Press, 2016.

[32] T. Mitsumata, S. Ohori, A. Honda and M. Kawai, " Magnetism and viscoelasticity of magnetic elastomers with wide range modulation of dynamic modulus," *Soft Matter,* vol. 9, no. 3, pp. 904-912, 2013.

[33] D. Romeis, V. Toshchevikov and M. Saphiannikova, "Elongated micro-structures in magneto-sensitive elastomers: a dipolar mean field model," *Soft Matter,* vol. 12, no. 46, pp. 9364-9376, 2016.

[34] A. A. Snarskii, M. Shamonin and P. Yuskevich, "Effective Medium Theory for the Elastic Properties of Composite Materials with Various Percolation Thresholds," *Materials,* vol. 13, 1243, 2020.

[35] A. V. Bodnaruk, A. Brunhuber, V. M. Kalita, M. M. Kulyk, P. Kurzweil, A. A. Snarskii, A. F. Lozenko, S. M. Ryabchenko and M. Shamonin, "Magnetic anisotropy in magnetoactive elastomers, enabled by matrix elasticity," *Polymer,* vol. 162, pp. 63-72, 2019.

[36] R. Hill, "A self-consistent mechanics of composite materials," *J. Mech. Phys. Solids,* vol. 13, no. 4, pp. 213-222, 1965.

[37] B. Budiansky, "On the elastic moduli of some heterogeneous materials," *J. Mech. Phys. Solids,* vol. 13, no. 4, pp. 223-227, 1965.

[38] A. K. Sarychev and A. P. Vinogradov, " Effective medium theory for the magnetoconductivity tensor of disordered material," *phys. stat. sol. (b),* vol. 117, no. 2, pp. K113-K118, 1983.

[39] D. T. Zimmerman, R. C. Bell, J. A. Filer, J. O. Karli and N. M. Wereley, "Elastic percolation transition in nanowire-based magnetorheological fluids," *Appl. Phys. Lett. ,* vol. 95, no. 1, 014102, 2009.

[40] R. W. Zimmerman, "Behavior of the Poisson ratio of a two-phase composite material in the high-concentration limit," *Appl. Mech. Rev.,* vol. 47, no. 1, pp. S38-S44, 1994.

[41] E. J. Garboczi and A. R. Day, "An algorithm for computing the effective linear elastic properties of heterogeneous materials: three-dimensional results for composites with equal phase Poisson ratios," *J. Mech. Phys. Solids,* vol. 43, no. 9, pp. 1349-1362, 1995.

[42] L. C. Davis, "Model of magnetorheological elastomers," *J. Appl. Phys. ,* vol. 85, no. 6, pp. 3348-3351, 1999.

[43] Z. Varga, G. Filipcsei and M. Zrínyi, "Magnetic field sensitive functional elastomers with tuneable elastic modulus," *Polymer,* vol. 47, no. 1, pp. 227-233, 2006.

[44] A. M. Menzel, "Mesoscopic characterization of magnetoelastic hybridmaterials: magnetic

gels and elastomers, their particle-scaledescription, and scale-bridging links," *Arch. Appl. Mech.,* vol. 89, no. 1, pp. 17-45, 2019.

[45] V. M. Kalita, Y. I. Dzhezherya and G. G. Levchenko, "Anomalous magnetorheological effect in unstructured magnetoisotropic magnetoactive elastomers.," *Appl. Phys. Lett.,* vol. 116, no. 6, 063701, 2020.

[46] D. Ivaneyko, V. Toshchevikov, M. Saphiannikova and G. Heinrich, " Effects of particle distribution on mechanical properties of magneto-sensitive elastomers in a homogeneous magnetic field," *Cond. Mat. Phys.,* vol. 15, no. 3, pp. 1-12, 2012.

[47] V. V. Sorokin, E. Ecker, G. V. Stepanov, M. Shamonin, G. J. Monkman, E. Y. Kramarenko and A. R. Khokhlov, "Experimental study of the magnetic field enhanced Payne effect in magnetorheological elastomers," *Soft Matter,* vol. 10, no. 43, pp. 8765-8776, 2014.

[48] V. V. Sorokin, I. A. Belyaeva, M. Shamonin and E. Y. Kramarenko, "Magnetorheological response of highly filled magnetoactive elastomers from perspective of mechanical energy density: Fractal aggregates above the nanometer scale?," *Phys. Rev. E,* vol. 95, no. 6, 062501, 2017.

**Acknowledgments:** We thank M. Snarskaya for Figure 1.

**Funding:** The research of M.S. was funded by the Deutsche Forschungsgemeinschaft (DFG, German Research Foundation), grant number 389008375.

**Author contributions:** Conceptualization, A.S. and M.S.; methodology, A.S.; formal analysis, A.S.; investigation, P.Y., A.S and M.S.; writing—original draft preparation, A.S. and M.S.; writing—review and editing, M.S. and A.S.; visualization, P.Y.; software, P.Y. and A.S.

**Competing interests:** Authors declare no competing interests.